\documentclass[sigconf,screen]{acmart}
\usepackage{multirow}
\usepackage{bm}
\usepackage[justification=centering]{caption}
\usepackage{balance}
\usepackage{float}
\usepackage{tabularx}
\usepackage{longtable}
\usepackage{hyperref}
\usepackage{xcolor}
\usepackage{colortbl}
\usepackage{fancyhdr}

\AtBeginDocument{%
  \providecommand\BibTeX{{%
    \normalfont B\kern-0.5em{\scshape i\kern-0.25em b}\kern-0.8em\TeX}}}

\setcopyright{acmcopyright}
\copyrightyear{2024}
\acmYear{2024}
\acmDOI{10.1145/xxxxxxx.xxxxxxx}
\acmConference[ASE '24]{ASE'24: the 39th IEEE/ACM International Conference on Automated Software Engineering (ASE '24)}{October 27 -- November 01, 2024}{Sacramento, California, USA}
\acmBooktitle{Proceedings of the 39th IEEE/ACM International Conference on Automated Software Engineering (ASE '24), October 27 -- November 01, 2024, Sacramento, California, USA}
\acmPrice{15.00}
\acmISBN{978-1-4503-XXXX-X/18/06}

\begin{document}
\begin{sloppypar}

\title{Copilot-in-the-Loop: Fixing Code Smells in Copilot-Generated Python Code using Copilot}

\author{Beiqi Zhang}
\affiliation{
\institution{School of Computer Science\\Wuhan University}
\city{Wuhan}
\country{China}
}
\email{zhangbeiqi@whu.edu.cn}

\author{Peng Liang}
\authornote{This work is partially funded by NSFC with No. 62172311.}
\affiliation{
\institution{School of Computer Science\\Wuhan University}
\city{Wuhan}
\country{China}
}
\email{liangp@whu.edu.cn}

\author{Qiong Feng}
\affiliation{
\institution{School of Computer Science\\Nanjing University of Science and Technology}
\city{Nanjing}
\country{China}
}
\email{qiongfeng@njust.edu.cn}

\author{Yujia Fu}
\affiliation{
\institution{School of Computer Science\\Wuhan University}
\city{Wuhan}
\country{China}
}
\email{yujia_fu@whu.edu.cn}

\author{Zengyang Li}
\affiliation{
\institution{School of Computer Science\\Central China Normal University}
\city{Wuhan}
\country{China}
}
\email{zengyangli@ccnu.edu.cn}

\renewcommand{\shortauthors}{B. Zhang et al.}


\begin{abstract}
As one of the most popular dynamic languages, Python experiences a decrease in readability and maintainability when code smells are present. Recent advancements in Large Language Models have sparked growing interest in AI-enabled tools for both code generation and refactoring. GitHub Copilot is one such tool that has gained widespread usage. Copilot Chat, released in September 2023, functions as an interactive tool aimed at facilitating natural language-powered coding. However, limited attention has been given to understanding code smells in Copilot-generated Python code and Copilot Chat's ability to fix the code smells. To this end, we built a dataset comprising 102 code smells in Copilot-generated Python code. Our aim is to first explore the occurrence of code smells in Copilot-generated Python code and then evaluate the effectiveness of Copilot Chat in fixing these code smells employing different prompts. The results show that 8 out of 10 types of code smells can be detected in Copilot-generated Python code, among which \textit{Multiply-Nested Container} is the most common one. For these code smells, Copilot Chat achieves a highest fixing rate of 87.1\%, showing promise in fixing Python code smells generated by Copilot itself. In addition, the effectiveness of Copilot Chat in fixing these smells can be improved by providing more detailed prompts. 
\end{abstract}



\maketitle

\section{Introduction} 
\label{Introduction}
Code smells refer to the symptoms of poor design and implementation decisions according to the definition by Martin Fowler in his book~\cite{fowler1999refactoring}. Code smells negatively affect the internal quality of software systems, hindering comprehensibility and maintainability \cite{palomba2018diffuseness, soh2016code} and increasing error proneness \cite{li2007empirical, olbrich2010are}. The identification of code smells suggests the potential need for code refactoring, pinpointing when and what refactoring can be applied to code \cite{fowler1999refactoring}.

Python, consistently ranked as one of the most popular programming languages \cite{TIOBE}, is increasingly used in various software development tasks. Python is a high-level, interpreted, and dynamic language that provides a simple but effective approach to object-oriented programming \cite{van1995python}. Due to Python's nature of flexibility and dynamism, developers often find it challenging both to write and maintain Python code \cite{chen2016detecting}, and abusing the short constructs of Python can expose code to bad patterns and reduce code readability \cite{lutz2010programming, beazley2009python}, leading to the occurrence of code smells in Python \cite{chen2018understanding}. 


Recent advancements in Large Language Models (LLMs) have unveiled impressive capabilities in solving various Natural Language Processing (NLP) tasks \cite{yang2024harnessing, wei2022chain}, showcasing their effectiveness in e.g., code generation \cite{izadi2024language} and refactoring\cite{sobania2023analysis, surameery2023use}. Released in June 2021, GitHub Copilot powered by LLM (i.e., OpenAI Codex) has been widely embraced by developers for code auto-completion, and it has evolved into the world's most widely adopted AI developer tool \cite{githubcopilot}. However, concerns arose regarding the quality of code generated by Copilot \cite{yetistiren2022assessing, zhang2023demystifying}. Copilot's code suggestion algorithms are incentivized to propose suggestions more likely to be accepted rather than easy to read and understand, which has an adverse impact for long-term code maintainability \cite{harding2024coding}. Subsequently, in September 2023, a public beta of GitHub Copilot Chat has been released as an interactive tool for Copilot, aiming to enable natural language-powered coding \cite{zhao2023copilotchat}. Developers can utilize Copilot Chat for tasks such as code analysis and fixing security issues, democratizing software development for a new generation. Given the potential for Copilot-generated Python code to exhibit code smells and the capacity of Copilot Chat to assist in rectifying such issues, this paper delves into fixing code smells in Copilot-generated Python code using Copilot Chat. \textbf{In this paper}, to evaluate the prevalence of code smells in Copilot-generated Python code and the competence of Copilot Chat in fixing Python smells, we built a dataset with 102 code smells in Copilot-generated Python code. Specifically, we investigated two Research Questions (RQs):

\begin{itemize}
    \item \textbf{RQ1}: To what extent does the Copilot-generated Python code contain code smells?
    \item \textbf{RQ2}: How effective is Copilot Chat in fixing different types of code smells in Copilot-generated Python code?
\end{itemize}



\textbf{Our preliminary findings} show that 14.8\% Python files generated by Copilot contain code smells, with \textit{Multiply-Nested Container} being the dominant code smell. 
GitHub Copilot exhibits promising potential in fixing code smells in Copilot-generated Python code, and the results indicate that Copilot Chat performs the best in fixing Python code smells by the prompt with code smell name. 

\section{Background \& Related Work} 
\label{Background & Related Work}

\subsection{Definition of Python Code Smells}
\label{Definition of Python code smells}
Considering the multiple programming paradigms and flexible grammatical constructs of Python (a dynamic programming language), the types of code smells presented by Martin Fowler \cite{fowler1999refactoring} that target in static programming language are not entirely applicable to Python code smells \cite{chen2016detecting}.
Chen \textit{et al.} \cite{chen2018understanding} proposed a set of 10 Python code smells (see Table \ref{definition of 10 python code smells}), which we considered in this study. These Python code smells are metric-based detectable using \texttt{Pysmell} \cite{chen2018understanding}, and these smells have been widely used in various studies (e.g., \cite{jebnoun2020scent, gesi2022code}) that explored code smells in Python projects.

\begin{table}[htbp]
\scriptsize
\caption{Definition of Python code smells.}
\label{definition of 10 python code smells}
\begin{tabular}{m{2.4cm}m{5.5cm}}
\toprule
\textbf{Code Smell}             & \textbf{Definition}  \\ \hline
Long Parameter List (LPL)       & A method or function with an extensive list of parameters \cite{fowler1999refactoring}.          \\ \hline
Long Method (LM)                & A method or function that exceeds a considerable length \cite{fowler1999refactoring}.            \\ \hline
Long Scope Chaining (LSC)       & A method or function with deeply nested closures \cite{fowler1999refactoring}.                   \\ \hline
Large Class (LC)                & A class that exceeds a considerable length \cite{fowler1999refactoring}.                         \\ \hline
Long Message Chain (LMC)        & An expression that accesses an object through an extended chain of attributes or methods using the dot operator \cite{brown1998antipatterns}. \\ \hline
Long Base Class List (LBCL)     & A class definition that inherits from an excessive number of base classes \cite{chen2018understanding}.      \\ \hline
Long Lambda Function (LLF)      & A lambda function that exceeds a considerable length \cite{chen2018understanding}.                           \\ \hline
Long Ternary Conditional Expression (LTCE) & A ternary conditional expression that exceeds a considerable length \cite{chen2018understanding}. \\ \hline
Complex Container Comprehension (CCC)  & A container comprehension (i.e., list, set, and dict comprehension, along with generator expression) with excessive complexity \cite{chen2018understanding}. \\ \hline
Multiply-Nested Container (MNC) & A container (i.e., set, list, tuple, and dict) with deep nesting \cite{chen2018understanding}. \\ \bottomrule
\end{tabular}
\end{table}

\subsection{Quality of Copilot-Generated Code}
Yetistiren \textit{et al.} \cite{yetistiren2022assessing} evaluated the quality of Copilot-generated code, focusing on validity, correctness, and efficiency. They found Copilot to be a promising tool for code generation tasks. Similarly, Nguyen and Nadi \cite{nguyen2022empricial} assessed Copilot-generated code for correctness and understandability. Their results show that Copilot-generated code exhibits low complexity, with no notable differences observed across programming languages. Pearce \textit{et al.} \cite{pearce2022asleep} investigated the prevalence and conditions that can cause GitHub Copilot to recommend insecure code. Their findings revealed that about 40\% of programs generated by Copilot were deemed vulnerable.

Different to the studies above, our work focuses on investigating a specific issue---code smells---in Copilot-generated Python code and accessing Copilot Chat's ability to fix these smells in the loop.

\section{Methodology} 
\label{Methodology}
\subsection{Use \texttt{Pysmell} to Detect Python Code Smells}
\label{use pysmell to detect python code smells}
This study aims to explore Python code smells in Copilot-generated code and evaluate Copilot Chat's capability to fix Python smells. We first built a dataset consisting of 102 code smells in Copilot-generated Python code by the following steps (see Figure \ref{fig: Overview of the dataset construction}):

\begin{figure}[htbp]
	\centering
	\includegraphics[width=1\linewidth]{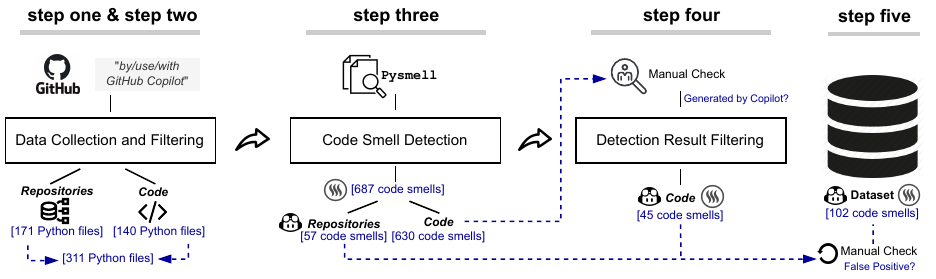}
	\caption{Overview of the dataset construction process.}
	\label{fig: Overview of the dataset construction}
\end{figure}

In \textbf{step one}, we used a keyword-based mining approach to collect Copilot-generated Python files from GitHub. Before the search process, a pilot search was conducted using the keyword ``\textit{GitHub Copilot}'' directly within GitHub's search engine. The pilot search results returned by GitHub were grouped into two categories, i.e., projects containing the keyword labeled as \textit{Repositories} and code files containing the keyword labeled as \textit{Code}. 
Under the \textit{Repositories} label, some projects claim in their README files or project descriptions that they were entirely generated by Copilot. Similarly, under the \textit{Code} label, certain files contain Copilot-generated code, as indicated by comments within the code.
However, the pilot search using ``\textit{GitHub Copilot}'' included many irrelevant results.
Our observations from the pilot search showed that using ``\textit{by GitHub Copilot}'', ``\textit{use GitHub Copilot}'', and ``\textit{with GitHub Copilot}'' as keywords could improve the accuracy of search results. 
Hence, we established the aforementioned three keywords as our search terms. 
The search was conducted on November 30, 2023, and Table~\ref{Search terms used in GitHub} shows the number of retrieved repositories and code. However, a Python file under the \textit{Code} label might contain multiple sets of keywords, implying duplicates among the 2,917 Python files we collected. 
After deduplication, 1,204 distinct Python files under the \textit{Code} label were retained.

\begin{table}[htbp]
\caption{Search terms used in GitHub.}
\scriptsize
\label{Search terms used in GitHub}
\begin{tabular}{m{0.8cm}<{\centering}m{2.8cm}m{2cm}<{\centering}m{1.6cm}<{\centering}}
\toprule
\textbf{\#}                                & \textbf{Search Term}              & \textbf{\# Repositories} & \textbf{\# Code} \\ \midrule
\textbf{ST1}                               & ``\textit{by GitHub Copilot}''    & 33                       & 896              \\
\textbf{ST2}                               & ``\textit{use GitHub Copilot}''   & 52                       & 1,069            \\
\textbf{ST3}                               & ``\textit{with GitHub Copilot}''  & 68                       & 952              \\ \hline
\specialrule{0em}{1pt}{1pt} \textbf{Total} &                                   & 153                      & 2,917            \\ \bottomrule
\end{tabular}
\end{table}


In \textbf{step two}, to manually label whether the projects under the \textit{Repositories} label were entirely generated by Copilot and whether the Python files under the \textit{Code} label contain Copilot-generated code, the first and fourth authors conducted a pilot data labelling by randomly selecting 10 candidates under each label. The two authors independently labelled whether these projects and code files should be included based on project documentation, code comments, and other metadata in the search results. Data labelled by the authors were compared, and the level of agreement between them were calculated using the Cohen's Kappa coefficient \cite{cohen1960coefficient}. The Cohen's Kappa coefficient was 0.79 for the projects under the \textit{Repositories} label and 0.85 for the code files under the \textit{Code} label, indicating a high level of agreement between the two authors. After the pilot labelling, the two authors checked all the projects and code files retrieved from GitHub. In the labelling process, if the two authors were unsure about whether a project or code file should be included, they discussed it with the second author until all three reached a consensus.
After manually filtering all the candidates, we collected 51 projects from \textit{Repositories} and 140 Python files from \textit{Code}. The 51 repositories comprised 171 Python files that were entirely generated by Copilot, while only a portion of the 140 code files were generated by Copilot. In total, we got 311 (171 + 140) Python files after this step (see Figure~\ref{fig: Overview of the dataset construction}). 
Note that the version information of Copilot that generated these Python files cannot be identified. 

In \textbf{step three}, we utilized \texttt{Pysmell}~\cite{chen2018understanding}, which has been widely used in various studies exploring code smells in Python \cite{jebnoun2020scent, vatanapakorn2022python, gesi2022code, sandouka2023python}, to detect the 10 code smells listed in Section \ref{Definition of Python code smells} in the 311 Python files obtained in \textbf{step two}. 
\texttt{Pysmell} has three thresholds for smell detection, and we opted for the Tuning Machine Strategy due to empirical evidence indicating its superior accuracy in detecting Python smells among the three strategies \cite{chen2018understanding}. A total of 687 code smells were detected by \texttt{Pysmell} in this step. All the code smells (57) detected in the Python files under the \textit{Repositories} label were generated by Copilot. However, under the \textit{Code} label, not all the detected Python smells (630) were Copilot-generated ones.

In \textbf{step four}, the first author manually checked whether the 630 instances of code smell under the \textit{Code} label obtained in \textbf{step three} were generated by Copilot. Only code smells in the Python files with code comments explicitly indicating that they were generated by Copilot were retained. During the manual checking process, in cases where the first author had uncertainties regarding the inclusion of a Python smell in the dataset, the second author was consulted for a discussion until an agreement was reached. Among the 630 code smells under the \textit{Code} label, 45 were generated by Copilot, resulting in a total of 102 (57 from \textit{Repositories} and 45 from \textit{Code}) code smells in Copilot-generated code were identified.

In \textbf{step five}, the first author rechecked all the identified code smells in \textbf{step four} to determine any potential false positives. After the manual check, all the 102 Python smells were confirmed as true positives, which was reasonable as \texttt{Pysmell} with Tuning Machine Strategy attains high precision in Python smell detection \cite{chen2018understanding}.

\subsection{Use Copilot Chat to Fix Python Code Smells}

\setcounter{table}{3}
\begin{table*}[htbp]
\scriptsize
\caption{Fixing rates for different types of code smells using different prompts.}
\label{tbl:fixing_rate_for_smell}
\begin{tabular}{m{2.6cm}m{1.3cm}<{\centering}m{1.3cm}<{\centering}m{1.3cm}<{\centering}m{1.3cm}<{\centering}m{1.3cm}<{\centering}m{1.3cm}<{\centering}m{1.3cm}<{\centering}m{1.3cm}<{\centering}m{1.3cm}<{\centering}}
\toprule
\textbf{Prompt}                          & \textbf{MNC}                        & \textbf{LPL}    & \textbf{LM}       & \textbf{LLF}                         & \textbf{LTCE}                        & \textbf{CC}                          & \textbf{CMC}     & \textbf{LC}      & \textbf{Avg}   \\ \hline
General Fix Prompt                       & \cellcolor{gray!5} 19.4\%          & \cellcolor{gray!0} 0.0\%           & \cellcolor{gray!15} 50.0\%           & \cellcolor{gray!45} \textbf{100.0\%} & \cellcolor{gray!30} 80.0\%           & \cellcolor{gray!15} 50.0\%           & \cellcolor{gray!15} 50.0\%         & \cellcolor{gray!45} \textbf{100.0\%} & 34.4\%          \\ \hline
Code Smell Fix Prompt                    & \cellcolor{gray!25} 58.3\%          & \cellcolor{gray!10} 22.7\%          & \cellcolor{gray!50} 91.7\%           & \cellcolor{gray!45} \textbf{100.0\%} & \cellcolor{gray!45} \textbf{100.0\%} & \cellcolor{gray!45} \textbf{100.0\%} & \cellcolor{gray!45} \textbf{100.0\%} & \cellcolor{gray!45} \textbf{100.0\%} & 64.5\%          \\ \hline
Specific Code Smell Fix Prompt           & \cellcolor{gray!25} \textbf{69.4\%} & \cellcolor{gray!40} \textbf{95.5\%} & \cellcolor{gray!45} \textbf{100.0\%} & \cellcolor{gray!45} \textbf{100.0\%} & \cellcolor{gray!45} \textbf{100.0\%} & \cellcolor{gray!45} \textbf{100.0\%} & \cellcolor{gray!45} \textbf{100.0\%} & \cellcolor{gray!45} \textbf{100.0\%} & \textbf{87.1\%} \\ \hline
\specialrule{0em}{1pt}{1pt} \textbf{Avg} & 49.1\%          & 39.4\%          & 80.6\%           & \textbf{100.0\%} & 93.3\%           & 83.3\%           & 83.3\%           & \textbf{100.0\%} &                 \\ \bottomrule
\end{tabular}
\end{table*}

\subsubsection{Prompt Design}
\label{prompt design}
Referring to the foundational prompt that instructs Copilot Chat to improve non-functional requirement of accessibility \cite{summers2023prompt}, we initially conducted a series of pilot experiments employing different prompt structures and formulations to fix code smells using Copilot Chat. Based on our prior observations, we selected three prompts of varying detail levels that demonstrated various effectiveness in fixing Python code smells:




\noindent \textbf{General Fix Prompt}: \textit{Fix the \textbf{problem} in the selected code}

\noindent \textbf{Code Smell Fix Prompt}: \textit{Fix the \textbf{code smell} in the selected code}

\noindent \textbf{Specific Code Smell Fix Prompt}: \textit{Fix the \textbf{[code smell name] (e.g., \textit{Long Method}) code smell} in the selected code}

\begin{figure}[htbp]
	\centering
	\includegraphics[width=0.72\linewidth]{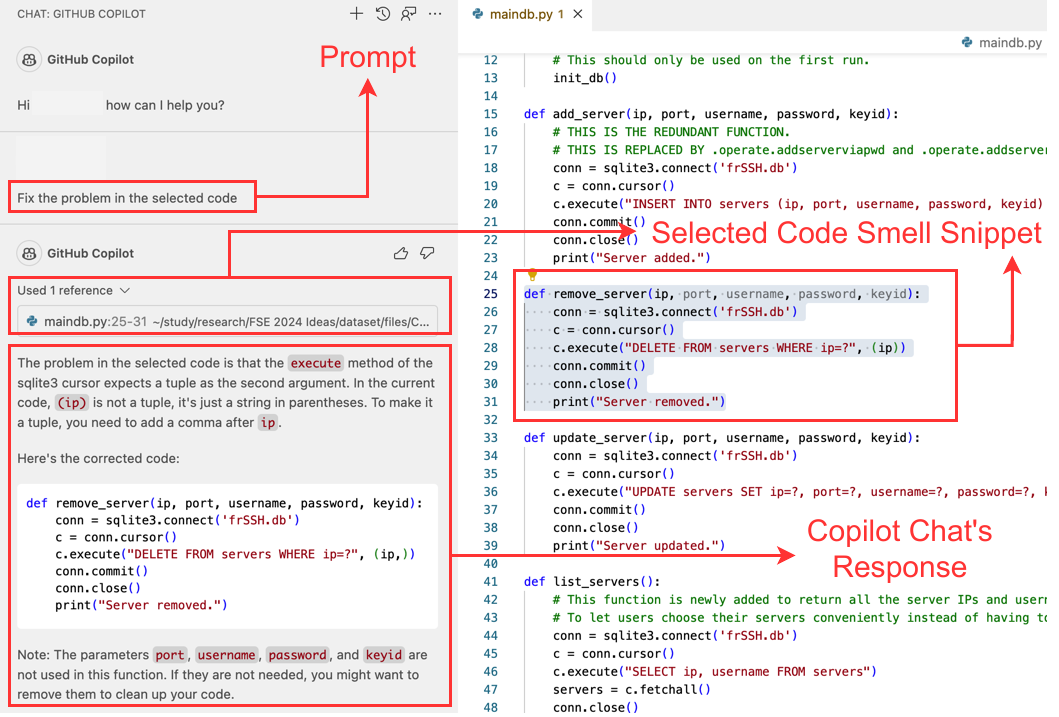}
	\caption{An example of using Copilot Chat to fix Python code smells in Visual Studio Code.}
	\label{fig: An example of using GitHub Copilot Chat to fix Python smells}
\end{figure}

The three prompts mentioned above each provide more details than the one before it, which enables us to explore Copilot Chat's effectiveness in fixing Copilot-generated code smells when provided with different levels of information. We first selected the code snippets of identified code smells 
and provided these 3 types of prompts to Copilot Chat in the chat window of Visual Studio Code. Copilot Chat then utilized the selected code as references (input) to generate responses to our prompts (see Figure~\ref{fig: An example of using GitHub Copilot Chat to fix Python smells}). 

\subsubsection{Code Smell Snippets Used as Referenced Code}
\label{code smell snippets used as referenced code}
\texttt{Pysmell} only provided the starting line and type of the code smells detected. We combined the information to determine the corresponding code snippet for each instance of the identified code smells. Consequently, 102 code smell snippets were obtained in Copilot-generated Python code at the class, method, or expression level. 
Notably, we observed that an individual code snippet could manifest multiple identical smells. To guide Copilot Chat in fixing code smell snippets, we constrained the selected snippets used as references to include at least one complete line of code, thereby ensuring that ample contextual information was provided to Copilot Chat. 
Hence, the number of code snippets that contain code smells for Copilot Chat to fix was reduced to 96. Among these, 3 exceeded the token length limit and were removed.
Out of the remaining 93 instances of code smells, 91 were in separate code snippets, while 2 different code smells were located in the same code snippet. We used that particular code snippet as a reference to address the 2 code smells, resulting in 92 code snippets encompassing the 93 distinct instances of code smells.
We used these 92 code smell snippets as referenced code and applied the prompts outlined in Section \ref{prompt design} as input to instruct Copilot Chat in fixing the smells. We recorded Copilot Chat's responses to our input for further evaluation, which are provided at \cite{replpack}.

\subsection{Evaluation of Code Smell Fixing}
To evaluate the effectiveness of Copilot Chat in fixing Python code smells generated by Copilot itself, the first author manually reviewed the code refactored by Copilot Chat utilizing the threshold in the Tuning Machine Strategy of \texttt{Pysmell} \cite{chen2018understanding} as the benchmark to verify if the code smell was fixed or not.
If the original code smell was resolved, we labeled it as ``\textit{Fixed}'', otherwise, we labelled it as ``\textit{Unfixed}''. We defined ``\textit{Fixing Rate}'' indicating the proportion of fixed smells 
to evaluate the effectiveness of code smell fixing.


\setcounter{table}{2}
\begin{table}[htbp]
\centering
\scriptsize
\caption{Code smell types detected in Copilot-generated Python code.}
\label{Types of code smells detected in Copilot-generated Python code}
\begin{tabular}{m{2cm}m{0.6cm}<{\centering}m{0.6cm}<{\centering}|m{2cm}m{0.6cm}<{\centering}m{0.6cm}<{\centering}}
\toprule
\textbf{Code Smell} & \textbf{\#}    & \textbf{\%}          & \textbf{Code Smell} & \textbf{\#}    & \textbf{\%}          \\ \hline
MNC                 & 41             & 40.2\%               & LTCE                & 5              & 4.9\%                \\ \hline
LPL                 & 22             & 21.5\%               & CCC                 & 4              & 3.9\%                \\ \hline
LM                  & 14             & 13.7\%               & LMC                 & 2              & 2.0\%                 \\ \hline
LLF                 & 12             & 11.8\%               & LC                  & 2              & 2.0\%                 \\ \bottomrule
\end{tabular}
\end{table}

\section{Results}
\label{Results}
\subsection{Results of RQ1}
\label{Results of RQ1}
\subsubsection{The proportion of Copilot-generated Python files that contain code smells}
In total, we collected 171 Python files from the \textit{Repositories} label and 140 from the \textit{Code} label.
Among these 311 (171+140) Python files, 46 contain code smells generated by Copilot, accounting for 14.8\%.

\subsubsection{The types of code smells detected in Copilot-generated Python code}
Table \ref{Types of code smells detected in Copilot-generated Python code} presents the 8 types of code smells detected in Copilot-generated Python code. Among the 10 detectable Python code smells listed in Section \ref{Definition of Python code smells}, 2 (LSC and LBCL) were not found in Copilot-generated code. \textit{MNC}, which accounts for over 40\%, is the most common type of code smell in Copilot-generated Python code, followed by \textit{LPL}, which represents over 20\%. \textit{LMC} and \textit{LC} are the least identified code smells in Copilot-generated Python code, each with a proportion of 2.0\% of the total.

\subsection{Results of RQ2}
\label{Results of RQ2}
Table \ref{tbl:fixing_rate_for_smell} lists Copilot Chat's average fixing rates with different prompts. 
Overall, the \textit{Specific Code Smell Fix Prompt}, which provides Copilot Chat with the particularized names of the code smells that needed to be fixed, achieved the highest average fixing rate at 87.1\%. On the other hand, the average fixing rate of the \textit{General Fix Prompt}, which instructs Copilot Chat to resolve potential issues in the referenced code without indicating the issue is a code smell, is the lowest (34.4\%). This result is in line with our intuition that using more detailed prompts to instruct Copilot Chat might get more effective code fix suggestions.

Copilot Chat's fixing rates for different types of code smells using different prompts are also showed in Table \ref{tbl:fixing_rate_for_smell}. In general, Copilot Chat demonstrates the best effectiveness in fixing \textit{LLF} and \textit{LC}, both achieving a fixing rate of 100.0\%. Conversely, Copilot Chat exhibits the lowest effectiveness in fixing \textit{LPL} and \textit{MNC}, with fixing rates of 39.4\% and 49.1\%, respectively. We can also find that when using the three prompts with varying levels of detail (see Section \ref{prompt design}) to instruct Copilot Chat in fixing the Python code smells detected in Copilot-generated code, the fixing rate with \textit{Specific Code Smell Fix Prompt} is consistently the highest for all the 8 types of code smells, while that with \textit{General Fix Prompt} is consistently the lowest.


\section{Discussion}
\label{Discussion}
\noindent \textbf{Attention to \textit{MNC} and \textit{LPL} in Copilot-generated Python code}: According to the RQ1 results (see Section \ref{Results of RQ1}), about 15\% Copilot-generated Python files contain code smells, and the top two code smells are \textit{MNC} and \textit{LPL}. However, based on the RQ2 results (see Section~\ref{Results of RQ2}), Copilot Chat shows the worst effectiveness in fixing these two code smells. The occurrence of \textit{MNC} reduces code readability and may obscure bugs, while \textit{LPL} makes code more complex \cite{chen2018understanding}, both negatively impacting the Python code quality. Therefore, developers should pay attention to \textit{MNC} and \textit{LPL} when using Copilot to generate Python code and Copilot Chat to fix them.

\noindent \textbf{Enhanced effectiveness through Detailed Prompts for Copilot Chat}: We used three prompts of varying detail levels to instruct Copilot Chat to fix code smells in Copilot-generated Python code. The RQ2 results (see Section \ref{Results of RQ2}) show that \textit{Specific Code Smell Fix Prompt}, providing comprehensive information, yielded the most favorable outcomes, while \textit{General Fix Prompt}, providing minimum information, produced the least effective results. This finding aligns with our expectation that Copilot Chat would exhibit better effectiveness in fixing Copilot-generated Python smells when offered more detailed prompts. When instructing Copilot Chat to fix Copilot-generated Python smells, developers can provide the specific type of code smell to improve Copilot Chat's effectiveness.



\section{Conclusions and Future Work}
\label{Conclusions and Future Work}
In this work, we constructed a dataset of 102 code smells in Copilot-generated Python code from GitHub, and evaluated the effectiveness of Copilot Chat in fixing these code smells.
The results show that 8 types of Python smell were detected in Copilot-generated code, and the dominant code smell is \textit{MNC}. Copilot Chat demonstrates promise in fixing Python smells in Copilot-generated code.

Potential avenues for future work include: (1) explore code smells in AI-generated code with other languages such as Java, C/C++ and Rust,
(2) investigate the impact of different prompt methods (e.g., few-shot learning \cite{brown2020language} and CoT prompting \cite{wei2022chain}) and LLM-based frameworks (e.g., RAG \cite{lewis2020rag} and multi-agent systems \cite{hong2023metagpt}) on Copilot Chat's ability to fix code smells,
and (3) explore the combination of Copilot Chat with other code analysis tools to enhance its code smell detection and fixing capabilities. 

Our work serves as \textbf{a starting point} for investigating the identification of code smells in AI-generated code and exploring the potential of AI-coding tools in fixing the smells by themselves.


\balance
\bibliographystyle{acm-reference-format}
\bibliography{reference}

\end{sloppypar}
\end{document}